\documentclass[
    ,final            
  ]
  {aipproc}

\layoutstyle{8x11single}

\begin{document}

\newcommand{\apjl}{Astrophysical Journal Letters}

\title{The Impact of Type Ia Supernova Ejecta on Binary Companions}


\author{P. M. Ricker}{
  address={Dept. of Astronomy, Univ. of Illinois, Urbana, IL 61801},
  email={pmricker@illinois.edu}
}

\author{K.-C. Pan}{
  address={Dept. of Astronomy, Univ. of Illinois, Urbana, IL 61801},
  email={kpan2@illinois.edu}
}

\author{R. E. Taam}{
  address={Dept. of Physics and Astronomy, Northwestern Univ., Evanston, IL 60208},
  altaddress={Academia Sinica Institute of Astronomy and Astrophysics, Taipei 10617, Taiwan},
  email={taam@tonic.astro.northwestern.edu}
}

\begin{abstract}
We present adaptive mesh refinement (AMR)
                        hydrodynamical simulations of the interaction
                        between Type Ia supernovae and their
                        companion stars within the context of the
                        single-degenerate model.  Results for 3D red-giant companions
                        without binary evolution agree with previous 2D
                        results by Marietta et al. We also consider
                        evolved helium-star companions in 2D.
                        For a range of helium-star
                        masses and initial binary separations, we
                        examine the mass unbound by the interaction
                        and the kick velocity delivered to the
                        companion star. We find that unbound mass versus
                        separation obeys a power law with index between -3.1 and -4.0,
                        consistent with previous results for
                        hydrogen-rich companions. Kick velocity also obeys a
                        power-law relationship with binary separation, but
                        the slope differs from those found for hydrogen-rich
                        companions. Assuming
                        accretion via Roche-lobe overflow, we find
                        that the unbound helium mass is consistent
                        with observational limits. Ablation (shock heating) appears
                        to be more important in
                        removing gas from helium-star companions than from
                        hydrogen-rich ones, though stripping (momentum transfer)
                        dominates in both cases.
\end{abstract}

\maketitle


\section{Introduction}

Type Ia supernovae (SNe~Ia), unlike core-collapse supernovae, are most likely purely
a phenomenon of close binary systems. Their defining absence of hydrogen
spectral lines and presence of silicon lines suggest an origin in
compact stellar remnants, while the progenitor delay times inferred from
their presence in early- and late-type galaxies and their absence from
interarm regions require these remnants to be low-mass, ie.\ white dwarfs (WDs).
The relative homogeneity of the SN~Ia population suggests that most white dwarfs
explode at a single mass, probably the Chandrasekhar mass, $1.44\ M_\odot$ \citep{whelan_binaries_1973}.
Because the distribution of white dwarfs peaks at $0.6\ M_\odot$ \citep{homeier_analysis_1998}, Chandrasekhar-mass
models require that the white dwarf accrete from a binary companion.

Most work on progenitor models
for SNe Ia has therefore focused on two general cases: the single-degenerate
model, in which a white dwarf accretes matter from a normal stellar binary companion, then
becomes unstable to explosive nuclear burning 
\citep{whelan_binaries_1973,nomoto_accreting_1982},
and the double-degenerate
model, in which two white dwarfs merge due to gravitational-wave emission
\citep{iben_supernovae_1984,webbink_double_1984}.
Single-degenerate models must cope with the problem of the companion star's hydrogen and helium; observational limits on H/He abundances in SNe~Ia are very constraining
\citep{mattila_early_2005,leonard_constrainingtype_2007}.
These models also have the difficulty that the mass accretion rate onto
the white dwarf must be in a fairly narrow range above $10^{-7}\ M_\odot\ {\rm yr}^{-1}$
to avoid either nova outbursts or development of a white-dwarf wind that limits the
accretion efficiency
\citep{nomoto_accreting_1982,hachisu_new_1996,ivanova_thermal_2004}.
This makes it difficult to explain the observed low-redshift SN~Ia rate of $3 \times 10^{-5}{\rm\ Mpc}^{-
3}{\rm\ yr}^{-1}$
\citep{mannucci_supernova_2005}
using single-degenerate progenitors alone.
The predicted and observed abundances of double-degenerate systems are sufficient
to explain the observed rate
\citep{nelemans_population_2001,napiwotzki_search_2001,napiwotzki_binaries_2002}.
However, such collisions may lead to production of an O-Ne-Mg white dwarf followed by
accretion-induced collapse to
a neutron star
\citep{nomoto_carbon_1985,ivanova_thermal_2004,dessart_multidimensional_2006,wickramasinghe_accretion_2009}.
It is possible that multiple progenitor channels contribute.
The observed redshift distribution of SNe~Ia is consistent with
at least two populations having different delay times measured from the zero-age
main sequence \citep{mannucci_supernova_2005}.

We are examining the single-degenerate model to determine, among other things, whether
the problem of H/He in the companion envelope can be overcome.
Single-degenerate progenitors can explode at the
Chandrasekhar mass for main-sequence (MS), red-giant (RG), or helium-star companions. If the
companion is a helium star, the WD may also rarely explode below the
Chandrasekhar mass via explosive burning of an accreted helium layer that
ignites interior carbon burning \citep{yungelson_supernova_2000}.
The Chandrasekhar-mass helium-star channel
is regarded as a candidate for the short delay-time population of SNe~Ia because
this case requires a massive initial MS star
\citep{kato_mass_2004,mannucci_two_2006,aubourg_evidence_2008,wang_helium_2009},
while the other two types of nondegenerate companion result from systems with low-mass stars and thus should produce longer delay times.

Multidimensional hydrodynamics simulations of the different channels are
difficult because of the large dynamic ranges required and
the large binary parameter space to explore. 
Simulations following the ejecta evolution within the progenitor system starting
from a pointlike explosion potentially make the most
direct contact with supernova observations and progenitor searches, though they
do not address the question of how the supernova explosion begins.
Such simulations can determine the amount of mass that should be
unbound from the companion, determine whether hydrogen lines should be seen,
determine the level of asymmetry introduced in the supernova remnant
 by interaction with the companion, and provide clues regarding the
appearance of the post-supernova companion.

Marietta et al.\ \cite{marietta_type_2000} explored the RG, subgiant, and MS single-degenerate channels
using 2D Eulerian hydrodynamics simulations. They found that
significant quantities of hydrogen should be unbound from the companion star
envelope in each case (15\% for MS and subgiant cases, and
98\% for the RG), in conflict with observational upper limits
\citep{mattila_early_2005,leonard_constrainingtype_2007}.
They also found that the companion star blocks about 7 - 12\% of the SN
ejecta, introducing anisotropy in the supernova remnant and possibly having
consequences for Si~II spectral line shapes, but that the kick delivered to the
companion is smaller than the orbital speed in each case and negligible for the
RG core. Fallback of part of the MS or subgiant companion's
 envelope produces a dramatic increase in the companion's luminosity,
but relatively little contamination by SN ejecta occurs. The remnant in
the RG case is a helium pre-white dwarf with an extended hydrogen-rich
envelope of very low mass (about $0.02\ M_\odot$) that should appear as an
underluminous O/B star for up to $10^6$~yr.

More recently, Pakmor et al.\ \cite{pakmor_impact_2008}
updated these results to consider the effect that pre-supernova
binary evolution has on the structure of MS companions, as suggested
by \cite{meng_impact_2007}, using 3D smoothed particle
hydrodynamics (SPH) simulations.
Pakmor et al.\ used initial conditions derived from stellar models 
\citep{ivanova_thermal_2004} that included binary mass transfer and thus yielded
a more compact MS companion. Because of this, they found a
tenfold reduction in the amount of unbound mass,
bringing the predicted amount back into agreement with observational upper
limits. They
also found strong dependences of the unbound mass $M_{\rm unbound}$
and kick velocity $v_{\rm kick}$ on initial binary separation $a$,
with $M_{\rm unbound} \propto a^{-3.49}$ and $v_{\rm kick} \propto a^{-1.45}$.
Their use of SPH with a relatively small number of particles
($\sim 10^6$) caused them to fail to reproduce the detailed fluid instabilities
seen by Marietta et al., but they performed a convergence study using the same
MS model as Marietta et al.\ and found good agreement 
for the unbound mass and kick velocity. These quantities therefore
should not be too sensitive to the method used, though we note that
the degree of contamination of the companion by SN ejecta should depend
significantly on the treatment of small-scale instabilities.

We are revisiting this problem using modern adaptive mesh refinement
(AMR) techniques, larger computers, and companion models that
incorporate the effects of binary evolution. These advances permit us to
consider larger 3D spatial dynamic ranges using the
same type of Eulerian shock-capturing hydrodynamics methods as 
Marietta et al. We also examine the helium-star channel, which
has not heretofore been considered using hydrodynamical simulations.
We have previously reported on 2D simulations of the helium-star channel
in \cite{2010ApJ...715...78P}; here we summarize these results and
present preliminary 3D results
for the RG channel.

\section{Numerical methods}

We use two separate codes to create progenitor stellar models
and to evolve SN explosions within binary systems. To construct 1D stellar models including mass loss due to binary
interactions, we use EZ 
\citep{paxton_ez_2004},
which is based on the STARS stellar evolution code originally developed by 
Eggleton \cite{eggleton_evolution_1971,eggleton_composition_1972}.
We have modified the code to include a mass loss term to account for binary mass
transfer and wind losses. Using parameter values chosen from 
\cite{ivanova_thermal_2004}
(for MS and RG companions) or
\cite{wang_helium_2009}
(for helium-star companions), we evolve each companion star model up to the point
at which the supernova explosion is expected to take place. We then interpolate the
1D profiles of density, temperature, and isotopic abundances onto the
2D or 3D grid used in the next stage of the calculations.
We allow the multidimensional stellar model to relax for
several dynamical times in order to reduce interpolation errors.

To simulate the explosion, we
employ FLASH 3.2
\citep{fryxell_flash:adaptive_2000,dubey_extensible_2009},
a parallel
Eulerian AMR hydrodynamical code based on the piecewise parabolic method or PPM
\citep{colella_piecewise_1984} for hydrodynamics and a direct multigrid algorithm
\citep{ricker_direct_2008} for gravity. 
We use a
nonideal equation of state (EOS) appropriate to stellar material that tabulates the Helmholtz
free energy as a function of density, temperature, and composition
\citep{timmes_accuracy_2000}.
We separately track hydrogen, helium, oxygen, and
carbon outside the SN ejecta, while the ejecta themselves are assigned a
pure nickel abundance.

FLASH uses the block-structured AMR package PARAMESH
\citep{macneice_paramesh:parallel_2000},
which manages an oct-tree mesh that is
distributed in parallel using a space-filling curve. To capture flow discontinuities,
we refine blocks based on the
second derivative of the gas density and pressure. We ignore the
second-derivative criterion in blocks where the maximum density falls below a preset
threshold ($10^{-6}$~g~cm$^{-3}$) in order to avoid excessively refining low-density
regions. To cope with features peculiar to each of our models, we employ additional
refinement criteria as described below.

Even with AMR, degenerate RG cores are too small for us to resolve
directly with our mesh while still achieving high resolution elsewhere in our
computational domain. Therefore we replace the core gas in RG companions
with a spherical cloud of about $2 \times 10^5$ particles with radius three
times the smallest zone spacing. The gravitational force on and due to the particles
is determined using the particle-mesh method
\citep{hockney_computer_1988}, with
an important modification: all of the particles in the cloud move rigidly together with the cloud's
center of mass. This arrangement ensures that the mapping of particle densities onto
the mesh and of gravitational forces onto the cloud's center of mass are free of
Cartesian grid effects introduced by the use of a cloud-in-cell (CIC) mapping kernel.
In hydrostatic equilibrium tests of a single RG star, this technique allows
us to maintain gas velocities in the envelope less than about 1\% of the sound speed
for more than a dynamical time. To ensure maximum mesh resolution of the
core, we force refinement of all mesh blocks that contain any particles in the cloud.
These methods have proven successful in related simulations of common-envelope
evolution 
\citep{ricker_interaction_2008}.
The MS and helium star
companions do not have degenerate cores, so we do not use particle clouds with them,
but since their outer density profiles are much steeper, we force refinement near their surfaces.

After allowing the companion star model to relax on a multidimensional AMR mesh,
we introduce an SN~Ia by adding a small spherical cloud of
high-energy gas. To reduce Cartesian mesh effects, we force the mesh to refine
about the explosion by
two extra levels beyond the rest of the computational volume within a spherical
region of radius five times the initial radius of the explosion. At this level of
refinement the initial explosion radius is about three zones across. For
Chandrasekhar-mass explosions, we use the W7 model
\citep{nomoto_accreting_1984} to
set the ejecta mass, kinetic energy, and thermal energy. The W7 model involves a
central carbon deflagration that provides a
good fit to observed SN~Ia light curves and spectra. In this model the explosion
energy is $1.233 \times 10^{51}$~erg, and the average radially-directed
ejecta velocity is
$8.527 \times 10^3$~km~s$^{-1}$. For our purposes, we assume the entire ejecta
mass ($1.378\ M_\odot$) consists of $^{56}$Ni.

\section{Results}

\begin{figure}
  \includegraphics*[height=.4\textheight,viewport=0 150 612 792]{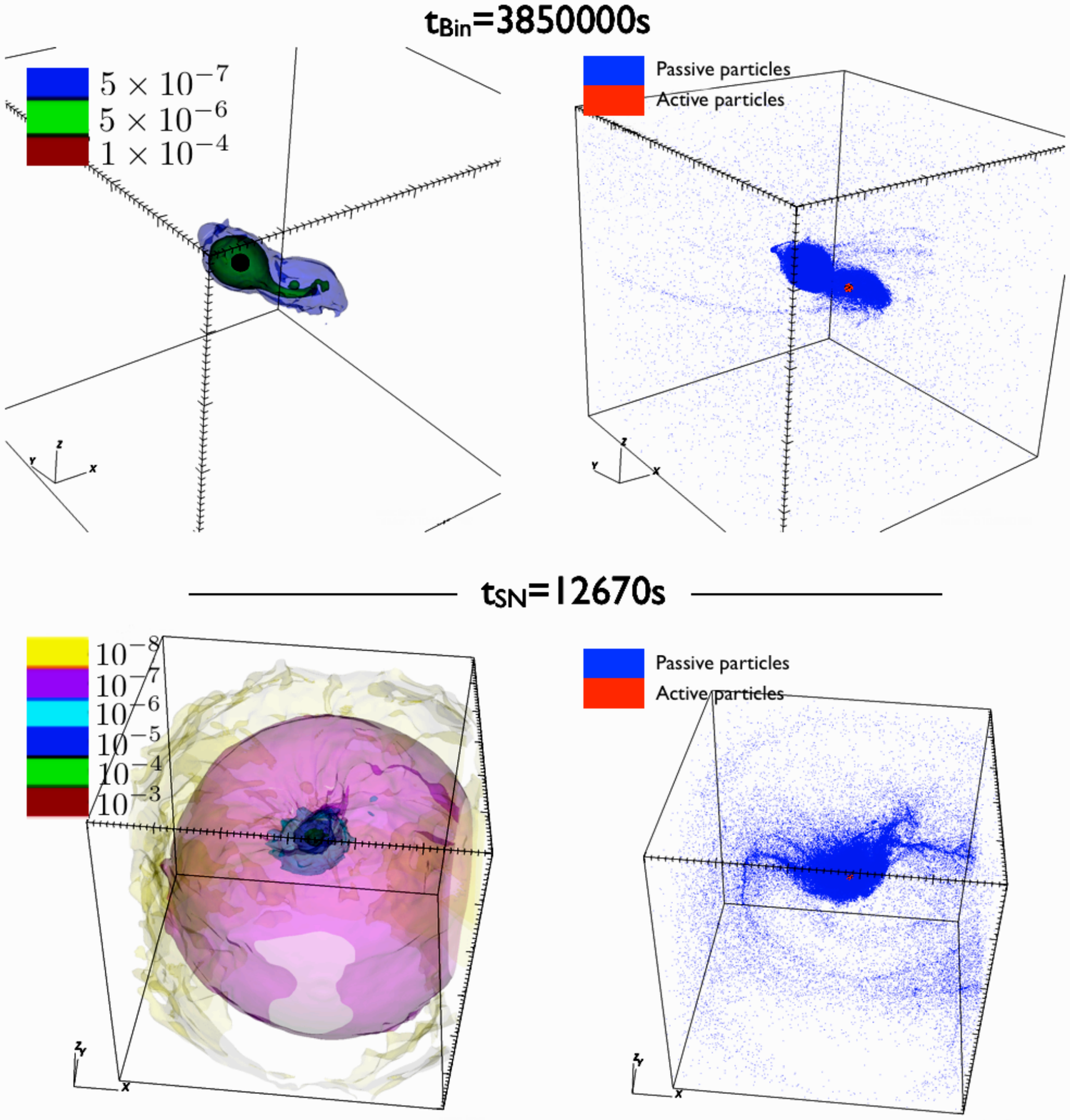}
  \ \ \ \ \ \ \ \ \ \ 
  \includegraphics*[height=.4\textheight,viewport=0 0 612 792]{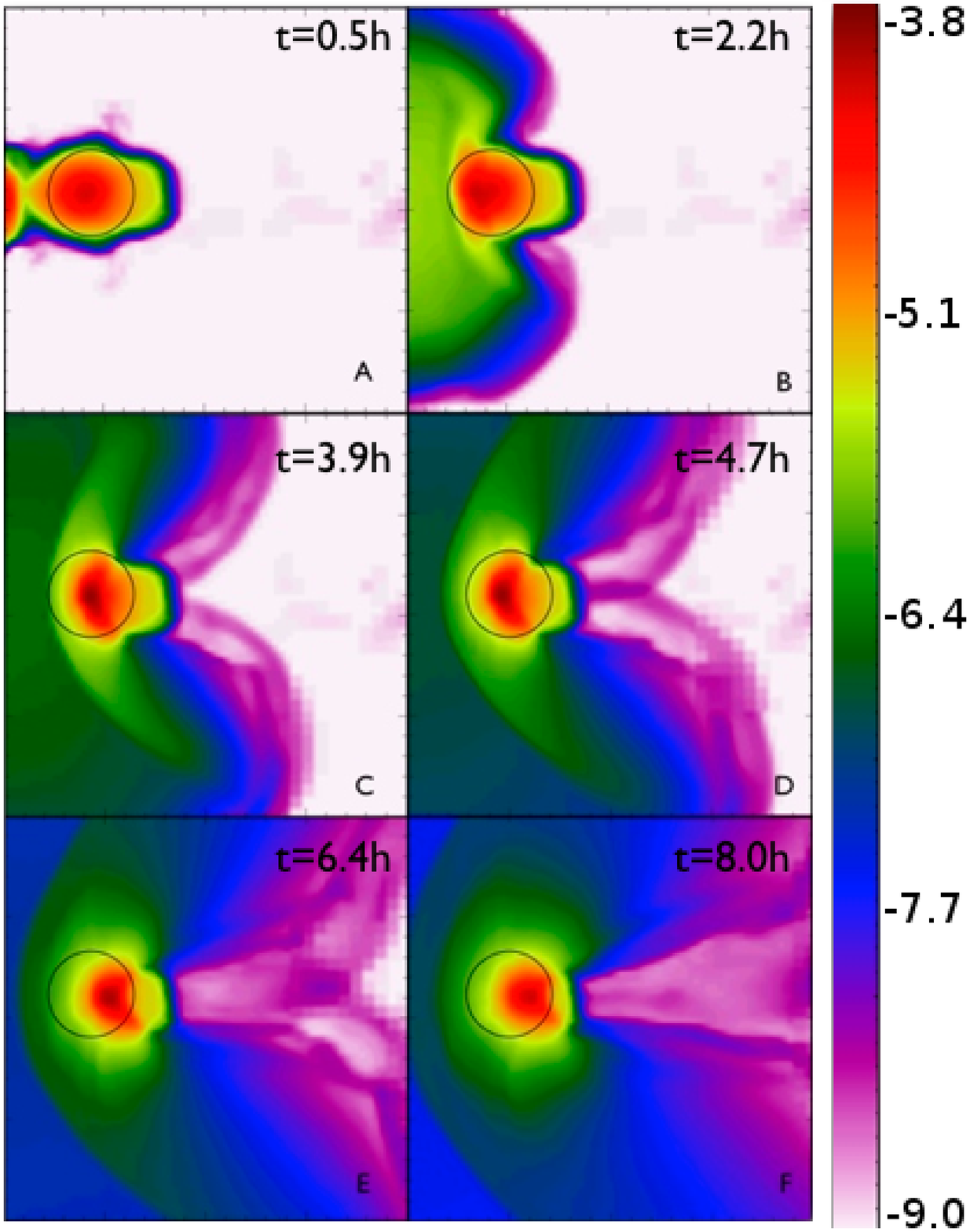}
  \caption{\label{Fig:3d rgwd}
           {\it Left panel:}
           Isodensity surfaces (left) and particle positions (right)
           for an RG-WD binary after $3.85\times 10^6$~s of binary
           evolution (top) and 12,670~s after a supernova explosion
           (bottom).
           Passive tracer particles are blue, while active particles
           representing the WD companion are red.
           The box size is about 2~AU ($3\times10^{13}$~cm).
           {\it Right panel:} Density distribution in the orbital plane for RG model
           at different times after a
           supernova explosion. The circle shows the initial RG size and position.
           Each frame spans a distance of 1~AU ($1.5\times 10^{13}$~cm). The
           color bar shows log density in g~cm$^{-3}$.
}
\end{figure}

We have created a preliminary 3D simulation of the RG case (Figure~\ref{Fig:3d rgwd})
for comparison with \cite{marietta_type_2000}.
The RG in this simulation had a $0.36\ M_\odot$
 core and a total mass of $1.05\ M_\odot$.  The initial binary separation was $6.2\times10^{12}$~cm,
and the minimum zone spacing was about $7.3\times10^{10}$~cm
within a computational volume of size 5~AU ($7.5\times10^{13}$~cm).
These initial conditions resulted in Roche-lobe overflow, so the system was followed for one
orbit with the unexploded WD represented by a particle cloud (as with the RG
core) to develop a realistic aspherical shape.
To initiate the explosion,
we removed the WD particle cloud and replaced it with
a gas sphere as described in the previous section.

Figure~\ref{Fig:3d rgwd} shows the double shock structure that forms in this simulation
when the SN ejecta begin to interact with the red giant.
The forward and reverse shocks are separated by a contact discontinuity.
When the reverse shock reaches the companion, a bow shock structure forms. We reproduce many features of
the RG companion case in \cite{marietta_type_2000},
but in our simulation an emptier central region is left
behind by the ejecta, because Marietta et al.\ imposed their explosion
 at the grid boundary, whereas ours is imposed inside the simulation
box. Moreover, because we include the orbital motion in 3D,
the ejecta become somewhat asymmetric in our calculation.
We use the kinetic, thermal, and potential energy of the gas in each zone to
determine the total amount of bound and unbound gas.
In agreement with \cite{marietta_type_2000}, we find that almost all the RG
envelope is unbound. (Note that neither simulation included the
effect of binary mass transfer on the initial companion model.)
However, a small amount of hydrogen is still left around the RG core.

Our investigation of the helium-star channel \citep{2010ApJ...715...78P}
has been more thorough
than for the RG case, but our simulations to date have been
2D and thus have not included orbital motion. We assumed
axisymmetry and placed the supernova and the helium star on the symmetry axis.
Because helium stars are much more compact than red giants, for these
simulations we used box sizes
of $1 - 5\times10^{11}$~cm and minimum zone spacings of $5\times 10^7$~cm to
$1.2\times10^8$~cm.

Qualitatively, our results are similar to the results of \cite{marietta_type_2000}
and 
\cite{pakmor_impact_2008}
for the MS case, but with a more compact
 companion and smaller binary separation. 
We performed a parameter survey using four different helium-star companions
suggested by \cite{wang_helium_2009},
described in Table~\ref{tab1}, and varying initial binary separations. 
The helium stars all have relatively low masses at the time of the supernova
explosion, but they evolve from MS stars of masses $5 - 8\ M_\odot$.
The white dwarfs in these models are the remnants of MS stars with initial
masses $2 - 6.5\ M_\odot$ and have accreted material up to the Chandrasekhar mass in the
course of the pre-supernova binary evolution.

\begin{figure}\includegraphics*[height=0.35\textheight,viewport=50 50 285 275]{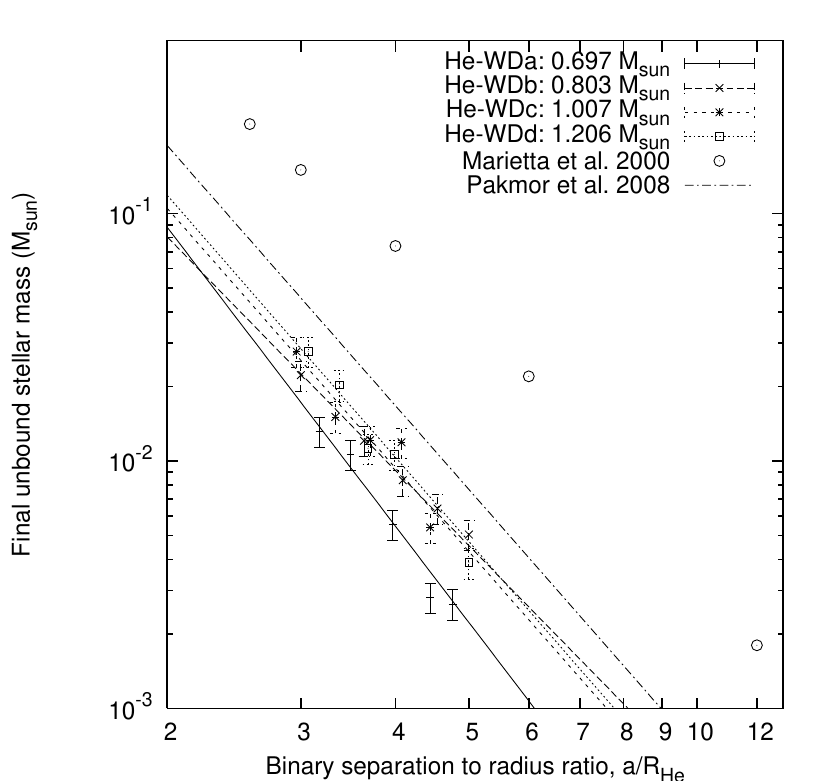}
            \includegraphics*[height=0.35\textheight,viewport=50 50 285 275]{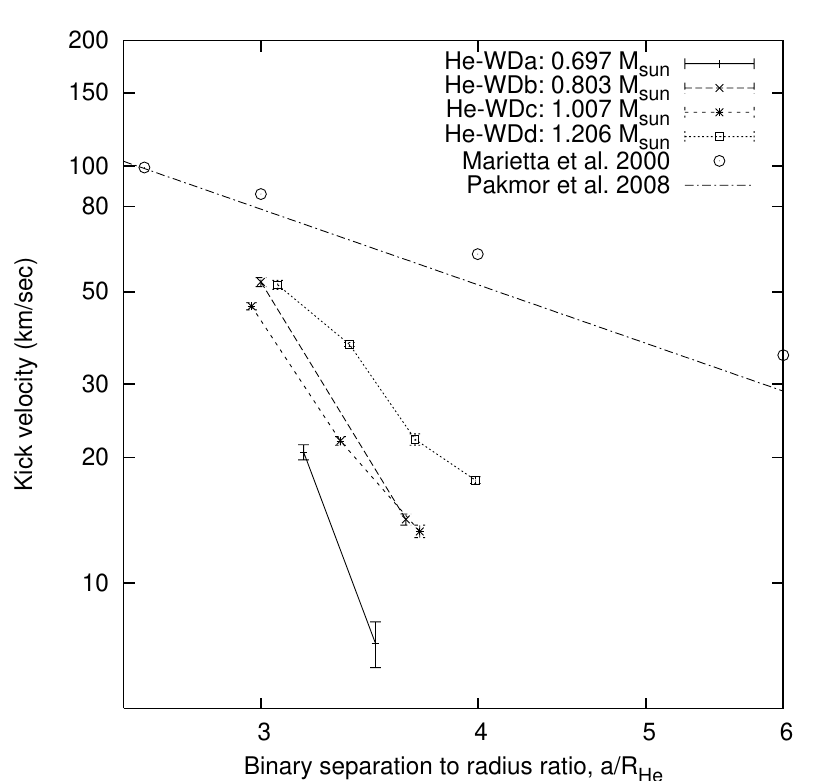}
\caption{\label{Fig:separation} Unbound mass and kick velocity versus initial binary separation for our different 2D helium-star models. Separation is scaled to the helium-star radius $R_{\rm He}$.}
\end{figure}

Figure~\ref{Fig:separation} shows the unbound mass (defined as total unbound helium)
and the helium-star kick velocity (defined as the center-of-mass velocity of the
bound helium) as functions of binary separation in our simulations. We include
14\% error bars based on results of a convergence test.
The unbound mass can be fit by the relation
\[
M_{\rm unbound}= C_{\rm ub} a^{m_{\rm ub}}\  M_\odot\ , \label{eq_stripped}
\]
where $a$ is the orbital separation, $m_{\rm ub}$ is the power-law index, and the constant $C_{\rm ub}$ depends only on the
helium-star model (see Table~\ref{tab1}).
For comparison, we also plot the power-law relation with index $-3.49$ found by
\cite{pakmor_impact_2008} and the data from \cite{marietta_type_2000} for MS
companions (consistent with an index of $-3.14$).
The power-law indices for our helium-star companions vary in a small range and
bracket their results, suggesting that the index may be insensitive to the evolutionary state
of the companion.
The normalization of the above relation does appear to be sensitive to the nature of the
companion star.

\begin{table}[b]
\begin{tabular}{ccccc}
\hline
            & Mass & Radius  & $m_{\rm ub}$ & $m_{\rm kick}$ \\
Model  & ($M_\odot$) & ($10^{10}$ cm) \\
\hline
He-WDa & 0.697 & 0.63 &  -4.01 & -3.28 \\
He-WDb & 0.803 & 1.10 &  -3.13 & -2.90 \\
He-WDc & 1.007 & 1.35 &  -3.48 & -3.18 \\
He-WDd & 1.206 & 1.63 &  -3.51 & -2.71 \\
\hline
\caption{ Helium-star models considered in \cite{2010ApJ...715...78P}
with best-fit power-law slopes.}\label{tab1}
\end{tabular}
\end{table}

At early times, the helium star velocity varies dramatically, but after about
1000 seconds it settles down to a roughly constant value. We use an appropriate
time average of this value as the kick velocity.
For initial binary separations larger
than 4 times the helium-star radius $R_{\rm He}$
the kick velocity could not be adequately determined because of the initial
potential perturbation of the WD. As obtained by \cite{pakmor_impact_2008} and \cite{meng_impact_2007},
a power-law relation is also found in our simulation and
can be fitted by the relation
\[
v_{\rm kick} = C_{\rm kick} a^{m_{\rm kick}} ,
\]
where $v_{\rm kick}$ is the kick velocity, $m_{\rm kick}$ is the power-law index, and the constant $C_{\rm kick}$ depends only on the helium-star model (see Table~\ref{tab1}). However, unlike the situation for the final unbound mass, the slope is very different from that found in the case of MS companions ($m_{\rm kick}  = -1.45$ in \cite{pakmor_impact_2008}, and
$m_{\rm kick} = -1.26$ in \cite{marietta_type_2000}). Based on a comparison with
the analytical model of \cite{meng_impact_2007}, we argue that this difference in
slopes may arise because ablation (shock heating) and stripping (momentum transfer)
have differing levels of importance in the helium-star case versus the RG and MS
cases. The relative contributions of these processes to mass loss from the companion
can be determined by examining the kinetic-to-thermal energy ratio of unbound
helium as in \citep{2010ApJ...715...78P} or, more accurately, by using passive tracer
particles (Figure~\ref{Fig:3d rgwd}) and determining this energy ratio at the
instant they become unbound.

The kick velocities for helium-star companions differ by a
factor of 2 or more from RG and MS companions
when considering separations characteristic of Roche-lobe
overflow ($a_{\rm RLOF} \approx 3 R_{\rm He}$ for the models considered here).
However, for helium stars the final velocity of the companion is dominated by the pre-SN
orbital velocity. Assuming the orbital velocity to be perpendicular to the kick,
we find that the kick velocity is only about 20\% of the resultant.

\section{Discussion and Conclusions}

While hydrogen-rich and helium-rich companion stars show many of the same qualitative
effects when struck by supernova ejecta, the degree of compactness of the companion
plays a very important role in determining how much mass is lost from the companion
and how much of a kick it receives. It remains to be seen whether the ratio of
ablated mass to stripped mass also depends on the degree of compactness, as this ratio has
not yet been computed for the RG and MS cases.
The companion's
compactness increases when we consider stars that
have already lost mass due to binary interactions. This suggests that an understanding
of the outcomes of common-envelope evolution is crucial for identifying the types of
companions most likely to encounter a single-degenerate SN~Ia situation.

Our next steps are to revisit the RG and MS channels via 3D AMR simulations that
include the effects of binary evolution and to examine the helium-star channel
in 3D. 3D simulations give us better estimates of the amount of SN ejecta
material that remains bound to the companion star, since the development of the
instabilities that drive mixing depends on grid dimensionality. They should also
permit us to determine more accurately the fraction of sightlines to the SN that
can be obscured by the companion.
 Using tracer particles in 3D, we can also more accurately determine
the relative amounts of stripped and ablated material (discussed using tracer
fluids in \citep{2010ApJ...715...78P}). Finally, by including radiation transport
we expect to determine what fraction of the hydrogen and helium unbound by the
supernova is ionized by radiation from the companion star.


\begin{theacknowledgments}
The simulations presented
here were carried out using the NSF Teragrid's Ranger
system at the Texas Advanced Computing Center under allocation TG-AST040034N. FLASH was developed
largely by the DOE-supported ASC/ Alliances Center for
Astrophysical Thermonuclear Flashes at the University
of Chicago. This work was supported, in part, by NSF
AST-0703950 to Northwestern University. PMR gratefully
acknowledges the International Travel Grant program of
the American Astronomical Society.
\end{theacknowledgments}



\bibliographystyle{aipproc}   

\bibliography{ms}

\begin{thebibliography}{37}
\expandafter\ifx\csname natexlab\endcsname\relax\def\natexlab#1{#1}\fi
\providecommand{\enquote}[1]{``#1''}
\expandafter\ifx\csname url\endcsname\relax
  \def\url#1{\texttt{#1}}\fi
\expandafter\ifx\csname urlprefix\endcsname\relax\def\urlprefix{URL }\fi
\providecommand{\eprint}[2][]{\url{#2}}

\bibitem[Whelan and Iben(1973)]{whelan_binaries_1973}
J.~Whelan, and I.~Iben, \emph{Astrophysical Journal} \textbf{186}, 1007--1014
  (1973).

\bibitem[Homeier et~al.(1998)]{homeier_analysis_1998}
D.~Homeier, D.~Koester, H.~Hagen, S.~Jordan, U.~Heber, D.~Engels, D.~Reimers,
  and S.~Dreizler, \emph{Astronomy and Astrophysics} \textbf{338}, 563--575
  (1998).

\bibitem[Nomoto(1982)]{nomoto_accreting_1982}
K.~Nomoto, \emph{Astrophysical Journal} \textbf{253}, 798--810 (1982).

\bibitem[Iben and Tutukov(1984)]{iben_supernovae_1984}
I.~Iben, and A.~V. Tutukov, \emph{Astrophysical Journal Supplement Series}
  \textbf{54}, 335--372 (1984).

\bibitem[Webbink(1984)]{webbink_double_1984}
R.~F. Webbink, \emph{Astrophysical Journal} \textbf{277}, 355--360 (1984).

\bibitem[Mattila et~al.(2005)]{mattila_early_2005}
S.~Mattila, P.~Lundqvist, J.~Sollerman, C.~Kozma, E.~Baron, C.~Fransson,
  B.~Leibundgut, and K.~Nomoto, \emph{Astronomy and Astrophysics} \textbf{443},
  649--662 (2005).

\bibitem[Leonard(2007)]{leonard_constrainingtype_2007}
D.~C. Leonard, \emph{Astrophysical Journal} \textbf{670}, 1275--1282 (2007).

\bibitem[Hachisu et~al.(1996)]{hachisu_new_1996}
I.~Hachisu, M.~Kato, and K.~Nomoto, \emph{Astrophysical Journal} \textbf{470},
  L97 (1996).

\bibitem[Ivanova and Taam(2004)]{ivanova_thermal_2004}
N.~Ivanova, and R.~E. Taam, \emph{Astrophysical Journal} \textbf{601},
  1058--1066 (2004).

\bibitem[Mannucci et~al.(2005)]{mannucci_supernova_2005}
F.~Mannucci, M.~D. Valle, N.~Panagia, E.~Cappellaro, G.~Cresci, R.~Maiolino,
  A.~Petrosian, and M.~Turatto, \emph{Astronomy and Astrophysics} \textbf{433},
  807--814 (2005).

\bibitem[Nelemans et~al.(2001)]{nelemans_population_2001}
G.~Nelemans, L.~R. Yungelson, S.~F.~P. Zwart, and F.~Verbunt, \emph{Astronomy
  and Astrophysics} \textbf{365}, 491--507 (2001).

\bibitem[Napiwotzki et~al.(2001)]{napiwotzki_search_2001}
R.~Napiwotzki, N.~Christlieb, H.~Drechsel, H.~Hagen, U.~Heber, D.~Homeier,
  C.~Karl, D.~Koester, B.~Leibundgut, T.~R. Marsh, S.~Moehler, G.~Nelemans,
  E.~Pauli, D.~Reimers, A.~Renzini, and L.~Yungelson, \emph{Astronomische
  Nachrichten} \textbf{322}, 411--418 (2001).

\bibitem[Napiwotzki et~al.(2002)]{napiwotzki_binaries_2002}
R.~Napiwotzki, D.~Koester, G.~Nelemans, L.~Yungelson, N.~Christlieb,
  A.~Renzini, D.~Reimers, H.~Drechsel, and B.~Leibundgut, \emph{Astronomy and
  Astrophysics} \textbf{386}, 957--963 (2002).

\bibitem[Nomoto and Iben(1985)]{nomoto_carbon_1985}
K.~Nomoto, and I.~Iben, \emph{Astrophysical Journal} \textbf{297}, 531--537
  (1985).

\bibitem[Dessart et~al.(2006)]{dessart_multidimensional_2006}
L.~Dessart, A.~Burrows, C.~D. Ott, E.~Livne, S.~Yoon, and N.~Langer,
  \emph{Astrophysical Journal} \textbf{644}, 1063--1084 (2006).

\bibitem[Wickramasinghe et~al.(2009)]{wickramasinghe_accretion_2009}
D.~T. Wickramasinghe, J.~R. Hurley, L.~Ferrario, C.~A. Tout, and P.~D. Kiel,
  \emph{Journal of Physics Conference Series} \textbf{172}, 2037 (2009).

\bibitem[Yungelson and Livio(2000)]{yungelson_supernova_2000}
L.~R. Yungelson, and M.~Livio, \emph{Astrophysical Journal} \textbf{528},
  108--117 (2000).

\bibitem[{Kato} and {Hachisu}(2004)]{kato_mass_2004}
M.~{Kato}, and I.~{Hachisu}, \emph{\apjl} \textbf{613}, L129--L132 (2004).

\bibitem[Mannucci et~al.(2006)]{mannucci_two_2006}
F.~Mannucci, M.~D. Valle, and N.~Panagia, \emph{Monthly Notices of the Royal
  Astronomical Society} \textbf{370}, 773--783 (2006).

\bibitem[Aubourg et~al.(2008)]{aubourg_evidence_2008}
{\'E.}.~Aubourg, R.~Tojeiro, R.~Jimenez, A.~Heavens, M.~A. Strauss, and D.~N.
  Spergel, \emph{Astronomy and Astrophysics} \textbf{492}, 631--636 (2008).

\bibitem[Wang et~al.(2009)]{wang_helium_2009}
B.~Wang, X.~Meng, X.~Chen, and Z.~Han, \emph{Monthly Notices of the Royal
  Astronomical Society} \textbf{395}, 847--854 (2009).

\bibitem[Marietta et~al.(2000)]{marietta_type_2000}
E.~Marietta, A.~Burrows, and B.~Fryxell, \emph{Astrophysical Journal Supplement
  Series} \textbf{128}, 615--650 (2000).

\bibitem[Pakmor et~al.(2008)]{pakmor_impact_2008}
R.~Pakmor, F.~K. R\"opke, A.~Weiss, and W.~Hillebrandt, \emph{Astronomy and
  Astrophysics} \textbf{489}, 943--951 (2008).

\bibitem[Meng et~al.(2007)]{meng_impact_2007}
X.~Meng, X.~Chen, and Z.~Han, \emph{Publications of the Astronomical Society of
  Japan} \textbf{59}, 835--840 (2007).

\bibitem[{Pan} et~al.(2010)]{2010ApJ...715...78P}
K.~{Pan}, P.~M. {Ricker}, and R.~E. {Taam}, \emph{Astrophysical Journal}
  \textbf{715}, 78--85 (2010).

\bibitem[Paxton(2004)]{paxton_ez_2004}
B.~Paxton, \emph{Publications of the Astronomical Society of the Pacific}
  \textbf{116}, 699--701 (2004).

\bibitem[Eggleton(1971)]{eggleton_evolution_1971}
P.~P. Eggleton, \emph{Monthly Notices of the Royal Astronomical Society}
  \textbf{151}, 351 (1971).

\bibitem[Eggleton(1972)]{eggleton_composition_1972}
P.~P. Eggleton, \emph{Monthly Notices of the Royal Astronomical Society}
  \textbf{156}, 361 (1972).

\bibitem[Fryxell et~al.(2000)]{fryxell_flash:adaptive_2000}
B.~Fryxell, K.~Olson, P.~Ricker, F.~X. Timmes, M.~Zingale, D.~Q. Lamb,
  P.~{MacNeice}, R.~Rosner, J.~W. Truran, and H.~Tufo, \emph{Astrophysical
  Journal Supplement Series} \textbf{131}, 273--334 (2000).

\bibitem[Dubey et~al.(2009)]{dubey_extensible_2009}
A.~Dubey, K.~Antypas, M.~K. Ganapathy, L.~B. Reid, K.~Riley, D.~Sheeler,
  A.~Siegel, and K.~Weide, \emph{Parallel Computing} \textbf{35}, 512--522
  (2009).

\bibitem[Colella and Woodward(1984)]{colella_piecewise_1984}
P.~Colella, and P.~R. Woodward, \emph{Journal of Computational Physics}
  \textbf{54}, 174--201 (1984).

\bibitem[Ricker(2008)]{ricker_direct_2008}
P.~M. Ricker, \emph{Astrophysical Journal Supplement Series} \textbf{176},
  293--300 (2008).

\bibitem[Timmes and Swesty(2000)]{timmes_accuracy_2000}
F.~X. Timmes, and F.~D. Swesty, \emph{Astrophysical Journal Supplement Series}
  \textbf{126}, 501--516 (2000).

\bibitem[{MacNeice} et~al.(2000)]{macneice_paramesh:parallel_2000}
P.~{MacNeice}, K.~M. Olson, C.~Mobarry, R.~de~Fainchtein, and C.~Packer,
  \emph{Computer Physics Communications} \textbf{126}, 330--354 (2000).

\bibitem[Hockney and Eastwood(1988)]{hockney_computer_1988}
R.~W. Hockney, and J.~W. Eastwood, \emph{Computer Simulation Using Particles},
  A. Hilger, Bristol {[England]}, 1988.

\bibitem[Ricker and Taam(2008)]{ricker_interaction_2008}
P.~M. Ricker, and R.~E. Taam, \emph{Astrophysical Journal} \textbf{672},
  {L41--L44} (2008).

\bibitem[Nomoto et~al.(1984)]{nomoto_accreting_1984}
K.~Nomoto, F.~Thielemann, and K.~Yokoi, \emph{Astrophysical Journal}
  \textbf{286}, 644--658 (1984).

\end{thebibliography}


\end{document}